\documentclass{emulateapj}

\newcommand{\eps}[1]{\mbox{log~$\epsilon$(#1)}} 

\newcommand\wave[1]{\mbox{$\lambda$#1\,\AA}}
\def\cs{\mbox{CS29497-030}}
\def\teff{\mbox{T$_{\rm eff}$}}
\def\logg{\mbox{log~{\it g}}}
\def\vmicro{\mbox{$\xi_{\rm t}$}}
\def\kmsec{\mbox{km~s$^{\rm -1}$}}

\shorttitle{BMP Star CS29497-030 in the Near-UV}
\shortauthors{Ivans et al.}

\begin{document}

\title{Near-UV Observations of CS29497-030:\\ New Constraints on Neutron-Capture Nucleosynthesis Processes}

\author{Inese I.\ Ivans\altaffilmark{2,3}, Christopher Sneden\altaffilmark{4},
Roberto  Gallino\altaffilmark{5}, \\ John J.\ Cowan\altaffilmark{6},
and George W.\ Preston\altaffilmark{7}}

\altaffiltext{2}{Dept.\ of Astronomy, California Institute of Technology,
	Pasadena, CA 91125; iii@astro.caltech.edu}

\altaffiltext{3}{Hubble Fellow.}

\altaffiltext{4}{Dept.\ of Astronomy, The University of Texas, 
	Austin, TX 78712; chris@verdi.as.utexas.edu}

\altaffiltext{5}{Dipartimento di Fisica Generale, Universita' di Torino,
	10125 Torino, Italy; gallino@ph.unito.it}

\altaffiltext{6}{Dept.\ of Physics and Astronomy, University of Oklahoma, 
	Norman, OK 73019; cowan@nhn.ou.edu}

\altaffiltext{7}{The Observatories of the Carnegie Institution of Washington, 
	Pasadena, CA 91101; gwp@ociw.edu}

\begin{abstract}
Employing spectra obtained with the new Keck~I HIRES near-UV 
sensitive detector, we have performed a comprehensive chemical 
composition analysis of the binary blue metal-poor star \cs.  
Abundances for 29 elements and upper limits for an additional seven 
have been derived, concentrating on elements largely produced via 
neutron-capture nucleosynthesis.  Included in our analysis are the 
two elements that define the termination point of the slow 
neutron-capture process, lead and bismuth.  We determine an
extremely high value of [Pb/Fe] = +3.65 $\pm$ 0.07 ($\sigma$ = 
0.13) from three features, supporting the single-feature result 
obtained in previous studies.  We detect Bi for the first time in a 
metal-poor star.  Our derived Bi/Pb ratio is in accord with those 
predicted from the most recent FRANEC calculations of the slow 
neutron-capture process in low-mass AGB stars.  We find that the 
neutron-capture elemental abundances of \cs\ are best explained by 
an AGB model that also includes very significant amounts of 
pre-enrichment of rapid neutron-capture process material in the 
protostellar cloud out of which the \cs\ binary system formed.  
Thus, \cs\ is both an $r$+$s$ and ``extrinsic AGB'' star.  
Furthermore, we find that the mass of the AGB model can be further 
constrained by the abundance of the light odd-element [Na/Fe] which 
is sensitive to the neutron excess.\footnote{The data presented herein were obtained at the W.\ M.\ Keck Observatory, which is operated as a scientific partnership among the California Institute of Technology, University of California, and NASA, and was made possible by the financial support of the W.\ M.\ Keck Foundation.}
\end{abstract}


\section{INTRODUCTION\label{intro}}

The bulk of the ``heavy elements'', those heavier than iron, are 
created by a combination of slow and rapid neutron-capture 
nucleosynthesis processes ($s$- and $r$-process) with each 
responsible for $\sim$50\% of the solar system 
isotopes.  In the $s$-process, successive neutron captures occur 
over sufficiently long timescales to permit unstable nuclei to 
$\beta$-decay and, in principle, the isotopic distribution of 
the $s$-process can be calculated from knowledge of stellar and 
nuclear physics (e.g., Busso, Gallino, \& Wasserburg 
1999\nocite{bgw99}; Straniero, Gallino, \& Cristallo 
2005\nocite{sgc05}). Pb and Bi are the last stable elements 
along the $s$-process path.  All isotopes heavier than Bi are 
unstable and $\alpha$-decay to Pb and Bi (Clayton \& Rassbach 
1967\nocite{cr67}).  In the Sun, the elemental abundances of 
Pb and Bi consist of significantly different combinations of 
$r$- and $s$-process isotope contributions, with $r$:$s$ 
ratios for Pb and Bi of 21:79 and 65:35, respectively 
(Travaglio et al.\ 1999\nocite{travaglio+99}; Simmerer et 
al.\ 2004\nocite{jensim+04}; and references therein).  The 
solar system chemical composition is the integrated result 
of many generations of stars, and depends upon the details 
of the formation history, initial mass functions, chemical 
yields, etc.  The Pb and Bi abundances most useful for 
unravelling the sites and nuclear parameters associated 
with the $s$- and $r$-process correspond to those in 
extremely metal-poor stars, formed from material with few 
prior generations of nucleosynthetic processing. 

In the last five years, dozens of low-metallicity stars with 
[Pb/Fe] $>$ 1 have been discovered\footnote{\footnotesize
We adopt the usual spectroscopic notation that for elements
A and B, \eps{A} $\equiv$ 
{\rm log}$_{\rm 10}$(N$_{\rm A}$/N$_{\rm H}$) + 12.0, and
[A/B] $\equiv$ 
{\rm log}$_{\rm 10}$(N$_{\rm A}$/N$_{\rm B}$)$_{\star}$ --
{\rm log}$_{\rm 10}$(N$_{\rm A}$/N$_{\rm B}$)$_{\odot}$.
e.g., [Pb/Fe] = 3 $\Rightarrow$ 
(N$_{\rm Pb}$/N$_{\rm Fe}$)$_{\star}$ = 
1000$\times$(N$_{\rm Pb}$/N$_{\rm Fe}$)$_{\odot}$.  Also, 
metallicity is defined as the stellar [Fe/H] value.}  (e.g.\ 
Sivarani et al.\ 2004\nocite{sivarani+04}; Barbuy et al.\ 
2005\nocite{barbuy+05}; and references therein).  Such large 
values of heavy $s$-process enhancements in low metallicity 
stars are thought to be the result of mass transfer in binary star
systems where the initially more massive star underwent an
asymptotic giant branch (AGB) evolutionary phase, and 
transferred material to the observed star.  The Pb 
enhancements observed in metal-poor  stars were predicted by 
\cite{gallino+98} who noted that lower metallicity stars were
expected to display increasingly higher abundances of 
heavier $s$-process elements relative to the abundances of 
lighter $s$-process elements.  At lower metallicities, the 
number of neutrons captured per iron seed increases, allowing 
heavier elements to be produced in greater abundance.  Pb 
enhancements have since been modelled by both phenomenological 
``classical'' AGB models with parameterized neutron exposures 
and ``stellar models'' with stellar astrophysical constraints 
(e.g.\ Arlandini et al.\ 1999\nocite{arlandini+99}; Delaude et 
al.\ 2004\nocite{delaude+04}; and references therein).  

Because Bi is the last stable element, knowledge of its
abundance in metal-poor halo stars will help pin down the 
predictions of the abundances of heavier radioactive 
actinide elements such as Th and U (Kratz et al.\ 
2004\nocite{kratz+04}; Ratzel et al.\ 2004\nocite{ratzel+04}; 
and references therein). Thus, abundance determinations of 
this element will also benefit nuclear chronometer studies of 
the age of the Galaxy.  With these multiple goals in mind, we 
observed the blue metal-poor star (BMP; Preston, Beers, \& 
Schectman 1994\nocite{pbs94}), \cs, a star which possesses 
the largest [Pb/Fe] abundance of any metal-poor star 
published to date (Sneden, Preston, \& Cowan 2003 -- 
SPC03\nocite{spc03}; Van Eck et al.\ 2003\nocite{van+03}, 
Sivarani et al.\ 2004\nocite{sivarani+04}; and references 
therein), in order to derive its abundance of Bi and other 
neutron-capture elements. 

\section{OBSERVATIONS, REDUCTIONS, AND ANALYSIS\label{obs}}

Spectra of \cs\ were obtained 2004 September 29--October 1
with the blue configuration of the Keck~I High Resolution 
Echelle Spectrometer (HIRESb; Vogt et al.\ 
1994\nocite{vogt+94}) and new 3-chip CCD mosaic 
(3$\times$2048$\times$4096$\times$15 $\mu$m pixels).  
The wavelength range coverage is essentially continuous in
the range $\sim$3050--5895 \AA.  With 3$\times$1 pixel 
on-chip binning and a slit width of 0.861 arcseconds, we 
obtained resolving power $R$ $\equiv$ 
$\lambda$/$\Delta\lambda$ $\simeq$ 40,000.  Nine 1800~s 
exposures were taken to attain a co-added signal-to-noise 
ratio (S/N) of 55:1 per resolution element in the centre 
of the blaze of the bluest order.  The S/N increases 
redwards, with $\sim$100 at \wave{3680}, $\sim$250 at 
\wave{4400}, and $\sim$300 at \wave{5850}.  Data reduction 
was performed using standard tasks in 
IRAF\footnote{\footnotesize
IRAF is distributed by NOAO, which is operated by AURA, 
under cooperative agreement with the NSF.}, 
FIGARO\footnote{\footnotesize 
FIGARO is provided by the Starlink Project which is run by 
CCLRC on behalf of PPARC (UK).}, and SPECTRE \citep{fs87}.

Our abundance analysis relied on the results of a combination 
of spectrum syntheses and equivalent width (EW) analyses 
measured with SPECTRE by fitting Gaussian profiles to the 
absorption lines. We employed stellar atmospheres without 
overshooting \citep{ck04}, and performed abundance 
calculations with a current version of MOOG \citep{sneden73}. 
Adopting the stellar parameters of SPC03 as initial parameters, 
we then iterated on the Fe abundances to eliminate abundance 
trends with respect to the excitation potentials, EWs, and 
ionization states.  We derive values of \teff\ = 7000~K, \logg\ 
= 4.1, \vmicro\ = 1.9 \kmsec, and [Fe/H] = --2.57, all in good 
agreement with SPC03\nocite{spc03}, with the exception of the 
lower metallicity derived from the higher quality data employed 
in this study.  We also compared our abundances for \cs\ to 
those of Sivarani et al.\ (2004)\nocite{sivarani+04}. With 19 
elements in common between the two studies (excluding those for
which only an upper limit was derived), most of the 
abundances agree to within 1-$\sigma$.  
Abundances in largest 
disagreement 
can be completely explained as the direct result 
of the differences in the adopted stellar parameters.
As Sivarani et al.\nocite{sivarani+04} show (their 
Table 2), the photometry of \cs\ leads to a wide range of 
\teff\ values based on colours.  In this study, we adopt 
the value of 7000~K, which satisfies both our spectroscopic 
constraints and is comparable to the estimate based on 
($V$--$K$) photometry.  Further technical details regarding 
our reduction and analysis techniques will be reported in an
expanded investigation currently underway (Ivans et al.\ 2005).

\begin{figure}
\epsscale{0.9}
\plotone{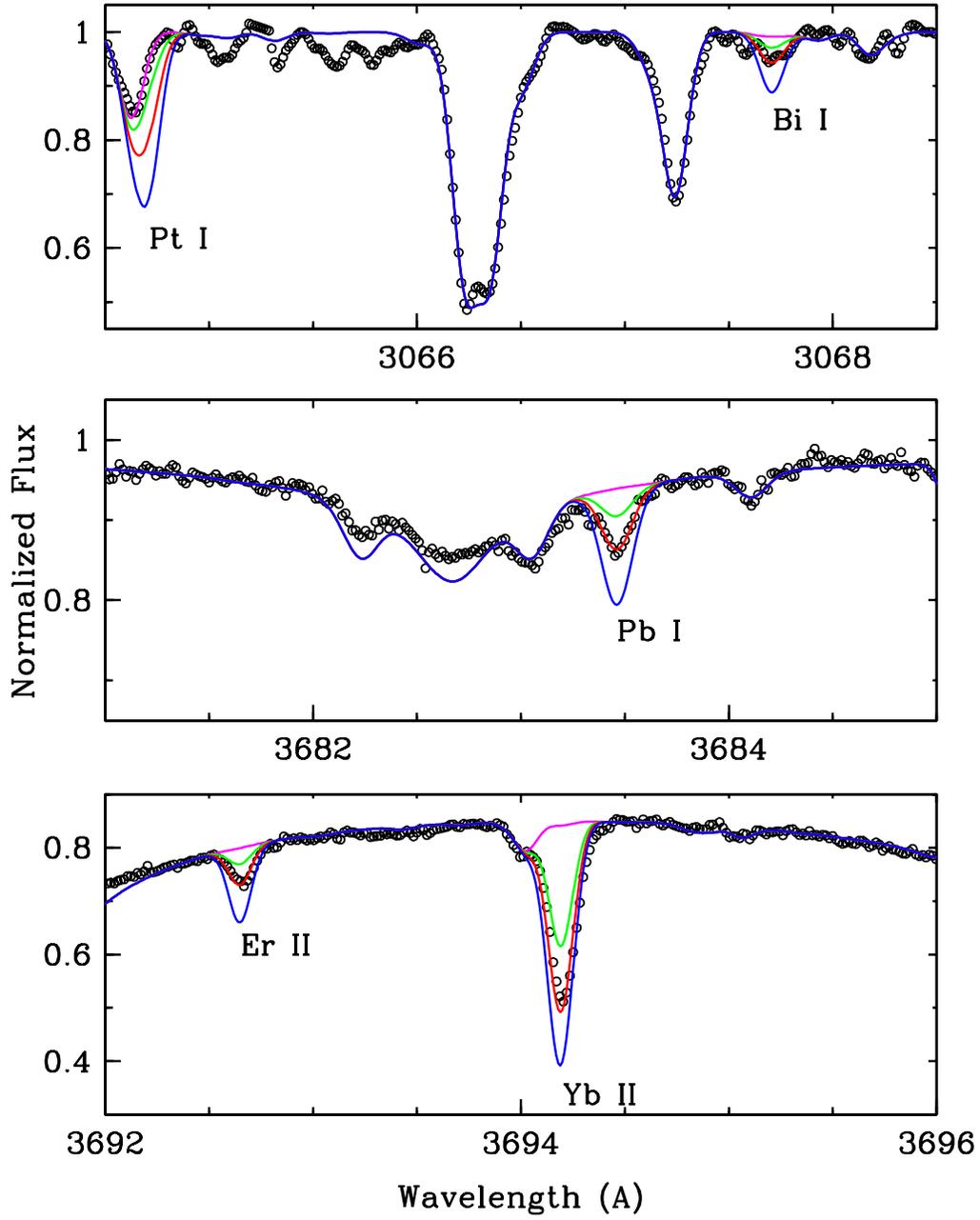}
\caption{
\label{cs.fig1}{\footnotesize
Spectra surrounding platinum, bismuth, lead, erbium, and 
ytterbium features in the near-UV.  The dots represent the 
observed spectrum and the solid lines, the spectrum 
syntheses.  The magenta line represents a synthesis with
no detectable contribution of the named element (e.g., 
only an upper limit is derived for platinum).  The green
and blue syntheses otherwise bracket the derived 
abundances (presented in Table~\ref{cs.tab1} and displayed 
here in red) by $\pm$ 0.4 dex.
}}
\end{figure}

In Figure~\ref{cs.fig1} we illustrate three 4\AA-swaths of 
spectrum syntheses surrounding regions of selected near-UV 
features of platinum, bismuth, lead, erbium, and ytterbium.  In 
Table~\ref{cs.tab1} we list our abundances for \cs, including 
formal uncertainties of their means ($\pm$), and the adopted 
1-$\sigma$, which includes an allowance for any uncertainty in 
the spectroscopically derived stellar parameters.  Abundances 
derived from a single feature have been assigned $\sigma$ = 
0.3~dex.  For some elements with additional useful features at 
wavelengths redward of our HIRESb set-up, we made new abundance 
determinations of the 2D-FRUTTI data employed 
by SPC03\nocite{spc03}.

The derived abundances of O, Na, Al, and K shown in 
Table~\ref{cs.tab1} have not been corrected for non-local
thermodynamic equilibrium (NLTE) effects which are known 
to exist for these elements.  The available literature on 
this issue in {\it warm} low metallicity stars is both 
sparse and in poor agreement.  For the elemental abundances of 
O, Na, Al, and K, we suggest that the following values be 
considered with greater confidence: upper limits of [O/Fe] 
$\leq$ +1.59; [Na/Fe] $\leq$ +0.88; [Al/Fe] $\leq$ +0.58; and 
[K/Fe] $\leq$ +1.07; which include our estimated NLTE 
corrections.  

\section{DISCUSSION\label{discuss}}

\begin{figure}
\epsscale{0.8}
\plotone{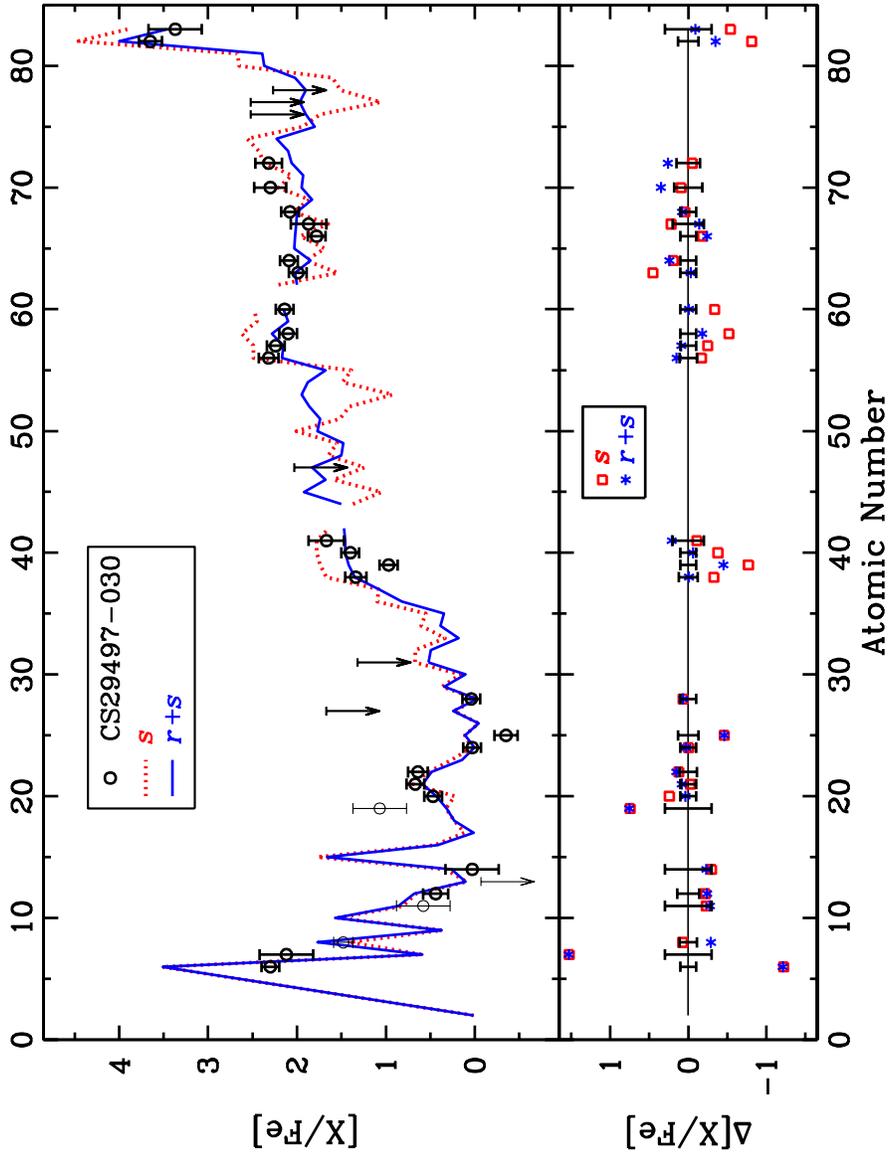}
\caption{
\label{cs.fig2}{\footnotesize
Comparison of the [X/Fe] abundances in CS29497-030 with 
predictions from $s$-process calculations of a 
1.3$M_{\sun}$ AGB star model.  In the top panel, the upper 
limits and open circles with error bars denote the 
stellar abundances.  Thinned symbols denote the four light 
element abundances which may suffer from uncorrected NLTE 
systematics (see \S~2).  The solid blue line represents 
the best fit $s$-process calculations based on an extreme 
$r$-process abundance pre-enrichment ($r$+$s$);  the red
dotted line represents predictions from $s$-process 
calculations without $r$-process enrichment.  The bottom 
panel displays the difference defined as $\Delta$[X/Fe] 
$\equiv$ [Fe/H]$_{\rm obs}$ -- [Fe/H]$_{\rm calc}$ and
upper limits are not shown.
}}
\end{figure}

We explored the origin of the neutron-capture elements 
in \cs\ by comparing the observed abundances with 
predicted $r$- and $s$-process contributions.  In 
Figure~\ref{cs.fig2}, we display our abundances in the 
context of $s$-process FRANEC calculations performed by 
Gallino et al.\  (1998\nocite{gallino+98}, 
2005\nocite{gallino+05}; also see Straniero et al.\ 
2005\nocite{sgc05}; and references therein).  Predictions 
from two sets of initial abundance assumptions are shown:
with and without pre-enrichment of the initial abundances.  
[La/Eu] ratios produced in the $s$-process at this low 
metallicity are typically +0.7~dex. In \cs\ the [La/Eu] 
ratio is $\sim$0.2, clearly indicating a strong 
$r$-process contribution. The pre-enrichment treatment 
employed in the $s$-process calculations here permits an 
exploration of the possibility that the initial abundances 
of \cs\ and its binary partner arose from a parent cloud 
with an extreme $r$-process abundance. In our picture, the 
formation of this pair of low mass stars was triggered by 
a supernova which polluted, snowplowed, and clumped a nearby 
molecular cloud.  In the associated $s$-process calculation, 
all of the initial $r$-process isotopes have been enhanced 
(according to their $r$-process contribution to the solar 
system abundances and normalized to Eu) and this, in turn, 
affects the seed abundances available to the subsequent 
$s$-processing.

In the case of no pre-enrichment, the displayed result
represents the most recent FRANEC calculations of the 
$s$-process at [Fe/H] $\sim$ --2.6 for an AGB star of 
1.3$M_{\sun}$ with an enhanced $^{13}$C abundance, in 
which all heavy-element abundance predictions are from the 
$s$-process.  Also illustrated in Figure~\ref{cs.fig2} 
are $s$-process calculations based on the $r$-process 
pre-enrichment scenario.  Calculations for $\pm$ 
0.05$M_{\sun}$ produce abundances which bracket 
those shown in the figure.  

The main component of the $s$-process is produced in an 
AGB star undergoing a series of He shell flashes via the 
triple-$\alpha$ reaction just below the H-burning shell.  
In these He-shell flashes (pulses), proton mixing leads to 
$^{12}$C(p,$\gamma$)$^{13}$N($e^{+}$,$\nu$)$^{13}$C($\alpha$,n)$^{16}$O 
reactions, releasing neutrons which can then be captured 
by Fe and heavy-element seeds (Iben \& Renzini 
1982\nocite{ir82}; Busso et al.\ 1999\nocite{bgw99}; 
Straniero et al.\ 2005\nocite{sgc05}).  The next thermal 
pulse injects the energy required to dredge up the 
nucleosynthesis products into the envelope while mixing 
more protons with the products of further triple-$\alpha$ 
reactions.  Thus, the photospheric abundance ratios of 
neutron-rich elements created in the $s$-process are a 
function of the histories of the envelope and core 
masses, and the number of thermal pulses.

The number of thermal pulses affects the number of free 
neutrons for subsequent neutron-capture processing and 
the abundance of Na places a stringent limit upon the 
assumed AGB star progenitor mass.  The Na abundance 
results from $^{22}$Ne(n,$\gamma$)$^{23}$Na reactions 
where $^{22}$Ne is largely of primary origin, only 
slightly affected by $\alpha$-captures during the thermal 
pulses.  $^{22}$Ne derives its abundance from CNO nuclei 
ashes ($^{14}$N) capturing $\alpha$-particles in the 
convective thermal pulse, with primary $^{12}$C produced 
and mixed to the surface by previous dredge-up episodes.  
More massive AGB models produce higher [Na/Fe] abundances 
(e.g., 1.5$M_{\sun}$ model undergoes $\sim$20 thermal 
pulses $\Rightarrow$ [Na/Fe] $\sim$2).  The best fit to 
our recommended upper limit of [Na/Fe], as well as the 
overall abundances of \cs, was found to be from a 
1.3$M_{\sun}$ AGB model that had undergone only six 
thermal pulses.

Among the light neutron-capture elements, where no
$r$-process enrichment was assumed, the $s$-process 
model with pre-enrichment predicts abundances in good 
agreement with those derived for Sr and Zr.  An 
``intrinsic AGB'' (i.e., high luminosity and low 
\logg) is expected to be Tc-rich, $^{93}$Zr-rich, and 
$^{93}$Nb-poor \citep{wd88}.  An ``extrinsic AGB'' is 
on or near the main sequence, and was once  the smaller 
mass star in a binary system.  The $s$-process 
abundances of the extrinsic AGB star are a result of 
pollution from the former AGB star's dredged-up 
material, at an epoch sufficiently remote for the 
$^{93}$Zr to have now decayed to $^{93}$Nb.  Our 
[Nb/Zr] ratio and stellar parameters for \cs\ are both 
in accord with those of an extrinsic AGB.  

The abundances of elements in \cs\ with large $s$-process 
contributions in the solar system are predicted to have 
lower abundances in the $r$+$s$ results than those predicted 
from the $s$-process operating without $r$-process 
pre-enrichment.  The pre-enrichment of the additional heavy 
elements affects the $s$-processing within the He-intershell 
of the AGB star, both in the $^{13}$C-pocket and the 
convective thermal pulse.  Because the heavy element 
isotopes in the pre-enrichment case are so numerous, they 
strongly compete with Fe as seeds for neutron-capture.  In 
addition, this competition takes a global role of leaving 
fewer neutrons available for the $s$-process isotopes which 
further decreases $s$-process efficiency.  

Other $r$+$s$ stars have recently been reported and
discussed in the literature (e.g.\ Aoki et al.\ 
2002\nocite{aoki+02}; Cohen et al.\ 2003; Johnson \&
Bolte 2004\nocite{jb04}; Zijlstra 2004; Barbuy et 
al.\ 2005\nocite{barbuy+05} and references therein).  In 
some of the previous efforts to model the abundances of 
$r$+$s$ stars, $s$-process material has been added to 
existing $r$-process abundance enhancements (e.g., 
Delaude 2004\nocite{delaude+04}).  However, those 
modelling attempts were performed employing fewer 
abundances than those employed here for \cs, and the 
previous attempts neither required nor included 
$s$-processing of the $r$-enhanced material within the 
He-intershell of the AGB star as has been done in the 
present study.  As noted, \cs\ possesses the largest 
[Pb/Fe] abundance of any metal-poor star published to 
date.  Its $s$-process contribution to Bi disguises its 
initial $r$-process contribution.  However, we predict 
that $r$+$s$ stars with {\em less} $s$-processing (and 
a relatively higher amount of $r$-processed material) 
will provide 3rd $s$-peak abundance ratios which can 
then be used to pin down the abundance predictions -- 
and production sites -- for the production of the 
$r$-process actinide elements such as Th and U.

\section{CONCLUSIONS\label{conclude}}

The chemical abundances of the extremely Pb-rich BMP star 
\cs\ are an excellent testbed to set new constraints on 
models of neutron-capture processes at low metallicity. 
We find \cs\ to be an $r$+$s$ star and that the abundance 
ratios are best fit by a pre-enrichment of $r$-process 
material out of which the binary system formed.  
The more massive companion underwent an AGB phase, and
heavy elements from the pre-enrichment competed with Fe 
as seeds for neutron-capture, leaving fewer neutrons
available for $s$-process isotopes, diminishing the 
$s$-process efficiency.  Pollution from the former AGB 
star's dredged-up material subsequently enriched the 
envelope composition of \cs.  Based on the fit to the 
low number of AGB model thermal pulses to match the 
observed abundance pattern, including the abundances of 
Na and Mg, we deduce that the progenitor mass of the AGB 
star was 1.3$M_{\sun}$.   The relative abundances we 
derive for Nb and Zr, as well as the stellar parameters 
corresponding to an evolutionary stage near the main 
sequence, confirm that \cs\ is an extrinsic AGB star.  

We encourage future studies of $s$-process abundance 
patterns to include the abundance of light elements 
sensitive to the neutron excess such as Na and Mg.  
We find that these elements are useful constraints on
the mass of the AGB progenitor.  However, some light 
elements are affected by NLTE effects and studies to 
date of those effects are sparse and/or in poor 
agreement.  It would be useful to extend the knowledge 
to lower metallicity and warmer temperatures than have 
so far been investigated.  

Overall, the predicted abundances from the $r$+$s$-process 
fit well the observed abundance patterns of \cs, from the 
first- (Sr and Zr) and second-peaks of the $s$-process (Ba
and La) as well as the third-peak (Pb and Bi).  Our abundance 
determination for Bi is the first such in any metal-poor star.  
And, the value derived for Bi is in accord with the Pb 
abundance.  We find that the $s$-process contribution to the 
pre-enrichment of $r$-process material in CS29497-030 swamps 
the $r$-process signature in the abundance of Bi.  We 
recommend that additional $r$+$s$ stars (but with relatively 
less $s$-processing than is found in \cs) be observed in order 
to derive Pb and Bi abundances that may further illuminate the 
issues regarding the production of actinide elements, Th and U 
and the other neutron-capture processes at work at early times 
in our Galaxy.

\acknowledgments %

We thank the following agencies for providing funding support for
this research: NASA through Hubble Fellowship grant HST-HF-01151.01-A 
from the Space Telescope Science Inst., operated by AURA, under NASA 
contract NAS5-26555 to III; NSF through grants AST03-07495 to CS and 
AST03-07279 to JJC; and MIUR-FIRB (Italy) ``The astrophysical origin 
of heavy elements beyond Fe'' to RG.  We are also grateful for the 
privilege of conducting observations from the revered summit of Mauna 
Kea.  We also acknowledge and appreciate the use of NASA's 
Astrophysics Data System Bibliographic Services; the successful 
efforts by the HIRES CCD upgrade group; the knowledgeable expertise of 
Keck staff during the run; and the generous help and patience of Sara 
Bisterzo in the generation of the FRANEC models.


\clearpage

\clearpage

\clearpage

\begin{center}
\begin{deluxetable}{lrrlrrr}
\tabletypesize{\scriptsize}
\tablenum{1}
\tablewidth{0pt}
\tablecaption{\cs: Derived Abundances\label{cs.tab1}}
\tablecolumns{7}
\tablehead{
\colhead{Species}          &
\colhead{Element}          &
\colhead{\eps{X}}          &
\colhead{$\pm$}            &
\colhead{$n$}              &
\colhead{[X/Fe]}           & 
\colhead{$\sigma$}         
}         
\startdata
C              & 6   & 8.29 & 0.03   &  9     & +2.30 & 0.10\\
C (CH)         & 6   & 8.46 &\nodata &\nodata & +2.47 & 0.10\\
N (CN)         & 7   & 7.60 &\nodata &\nodata & +2.12 & 0.35\\
O{\tablenotemark{(a)}}
               & 8   & 7.84 & 0.11   &  3     & +1.48 & 0.11\\ 
Na{\tablenotemark{(a)}}
               &11   & 4.34 &\nodata &  1     & +0.58 & 0.30\\
Mg             &12   & 5.45 & 0.07   &  4     & +0.44 & 0.14\\
Si             &14   & 5.01 &\nodata &  1     & +0.03 & 0.30\\
K{\tablenotemark{(a)}}
               &19   & 3.62 &\nodata &  1     & +1.07 & 0.30\\
Ca             &20   & 4.26 & 0.04   &  6     & +0.47 & 0.10\\
Sc             &21   & 1.20 & 0.01   &  3     & +0.67 & 0.10\\
Ti             &22   & 3.06 & 0.03   & 17     & +0.64 & 0.11\\
Cr             &24   & 3.13 & 0.04   &  5     & +0.03 & 0.10\\
Mn             &25   & 2.47 & 0.07   &  4     & -0.35 & 0.13\\
Ni             &28   & 3.72 & 0.02   &  2     & +0.04 & 0.10\\
Sr             &38   & 1.67 & 0.07   &  3     & +1.34 & 0.12\\
 Y             &39   & 0.64 & 0.02   & 13     & +0.97 & 0.10\\
Zr             &40   & 1.43 & 0.03   & 11     & +1.40 & 0.10\\
Nb             &41   & 0.52 & 0.04   &  3     & +1.67 & 0.20\\
Ba             &56   & 1.88 & 0.06   &  3     & +2.32 & 0.11\\
La             &57   & 0.80 & 0.01   & 15     & +2.22 & 0.10\\
Ce             &58   & 1.08 & 0.02   & 20     & +2.10 & 0.10\\
Nd             &60   & 1.02 & 0.02   & 17     & +2.14 & 0.10\\
Eu             &63 & --0.07 & 0.03   &  6     & +1.99 & 0.10\\
Gd             &64   & 0.64 & 0.02   &  5     & +2.09 & 0.10\\
Dy             &66   & 0.31 & 0.02   &  7     & +1.78 & 0.10\\
Ho             &67 & --0.19 & 0.05   &  4     & +1.87 & 0.20\\
Er             &68   & 0.44 & 0.04   &  4     & +2.08 & 0.10\\
Yb             &70   & 0.81 & 0.13   &  2     & +2.30 & 0.18\\
Hf             &72   & 0.63 & 0.07   &  5     & +2.32 & 0.15\\
Pb             &82   & 2.93 & 0.07   &  3     & +3.65 & 0.13\\
Bi             &83   & 1.51 &\nodata &  1     & +3.37 & 0.30\\
\cutinhead{Upper Limits}
Al{\tablenotemark{(a)}}
               &13   & 3.83 &\nodata& 1     & -0.07 &\nodata\\
Co             &24   & 4.02 &\nodata& 1     & +1.67 &\nodata\\
Ga             &31   & 1.63 &\nodata& 1     & +1.32 &\nodata\\
Ag             &47   & 0.7: &\nodata& 1     & +2.03 &\nodata\\
Os             &76   & 1.4: &\nodata& 1     & +2.52 &\nodata\\
Ir             &77   & 1.3: &\nodata& 1     & +2.52 &\nodata\\
Pt             &78   & 1.5: &\nodata& 1     & +2.27 &\nodata\\
\enddata
\small{
\tablenotetext{(a)}{Derived abundance does not take into 
account NLTE correction which would revise abundance.  
See \S~2 for discussion and recommended upper limits.}
}
\end{deluxetable}
\end{center}

\end{document}